\documentclass[twocolumn,showpacs,preprintnumbers,amsmath,amssymb]{revtex4}
\usepackage{graphicx}
\usepackage{dcolumn}
\usepackage{bm}
\usepackage{colortbl}

\begin{document}

\preprint{}

\title{Angular distributions of atomic photoelectrons produced in the UV and XUV regimes}

\author{S. Bauch}
\author{M. Bonitz}
 \email{bonitz@physik.uni-kiel.de}
\affiliation{%
Institut f\"ur Theoretische Physik und Astrophysik\\
Christian-Albrechts-Universit\"at zu Kiel, D-24098 Kiel, Germany
}%

\date{\today}

\begin{abstract}
We present angular distributions of photoelectrons of atomic model systems excited by intense linearly polarized laser pulses in the VUV- and XUV-regime. We solve the multi-dimensional time-dependent Schr\"odinger equation for one particle on large spatial grids and investigate the direction dependence of the ionized electrons for isotropic $s$-states as well as $p$-states. Although the ponderomotive potential is small compared to the binding energy of the initially bound electron and the photon energy of the exciting laser field, richly structured photoelectron angular distributions are found which sensitively depend on the laser frequency and intensity as well as on the number of absorbed photons. The occuring shapes are explained in terms of scattering mechanisms.
\end{abstract}

\maketitle

\begin{section}{Introduction}
\label{section:introduction}
Photoemission experiments are an important tool for the investigation of electronic properties of matter from single atoms and molecules to condensed matter systems, see e.g. \cite{schattke_hove} for an overview.

The first experimental investigation of photoelectron angular distributions (PADs) from atoms in the gas phase using ultra-violet light was published in 1930 \cite{PhysRev.36.6}. The shape of the PAD followed the $\cos^2$ dipole-like shape predicted by early quantum mechanics \cite{Sommerfeld.1919} which stated that the most favorable direction of emitted electrons is along the field polarization axis.\\
The technical development of new light sources, in the first place the invention of the laser \cite{Nat.187.493}, with the ability of creating high-intensity and monochromatic pulses gave new momentum to this topic, giving access to non-linear processes. An (ionization) process is called non-linear if the photon energy $\hbar \omega$ is smaller than the ionization potential $I_p$ of the atomic state considered. According to the classification by Keldysh \cite{JETP.20.1307} two main processes can be distinguished. Depending on the parameter $\gamma= \sqrt{I_p/2U_p}$ (where $U_p=e_0^2E_0^2/4 m_e \omega^2$ is the ponderomotive energy of the electron in the changing electric field of the laser with amplitude $E_0$) the electron is freed into a continuum state by tunnel ionization ($\gamma < 1$) or by multi-photon (MP) ionization ($\gamma > 1$).

After its ionization the electron can absorb additional photons if the intensity of the laser is sufficiently high. The first experimental evidence of such \emph {above-threshold ionization} (ATI) by one additional photon in intense infrared (IR) laser pulses has been observed in the late 70's \cite{PhysRevLett.42.1127}. In this case, the photoelectron is ionized by a MP or tunneling process and absorbs additional photons from the light field to increase its kinetic energy leading to a peak-like structure in the electron spectra, where each peak is separated by the photon energy $\hbar \omega$ of the exciting laser. Soon, better laser systems and higher accuracy in experiments showed the expected sequence of peaks in the continuum \cite{PhysRevA.28.248, JPhysB.14.L597} which form well-known plateau structures. 
The \emph {quasistatic} or \emph {two-step} model \cite{PhysRevLett.71.1994, PhysRevLett.62.1259,PhysRevLett.61.2304} explains the underlying dynamics. First, the electron is ejected into a continuum state via tunneling or MP processes with nearly no kinetic energy, being subsequently accelerated by the electrical field of the laser. The main structures of ATI spectra can be explained by rescattering processes where parts of the electronic wave function are driven back to the parent ion. This effect is the origin of such famous and actively studied processes as high harmonics generation (HHG), non-sequential double (or multi) ionization and the formation of the characteristic cut-offs in the energy distribution of ATI electrons.\\

A large amount of information regarding electron dynamics in atoms and ionization processes is obtained by analyzing the corresponding angular distributions of electrons. 
First investigations of multi-photon PADs revealed intensity-dependent structures \cite{PhysRevA.29.2677}. According to the simple quasistatic model the general thought was that the PADs are more and more peaked along the polarization axis of the laser field with increasing photon order of the process, which leads to occupation of final states with higher angular momentum. But the experimental situation soon changed with the appearence of additional structures referred to as \emph {side lobes}\cite{PhysRevLett.71.3770}, \emph {jets} and \emph {wings}\cite{JPhysB.31.4617} which appear at characteristic regions in the ATI photoelectron spectra, depending on the ponderomotive energy. 
The angular distributions have been widely used to identify the involved high-lying Rydberg states (``channel switching'') whose angular momentum significantly affects the final angular momentum state of the photoelectron, and such the PAD \cite{PhysRevA.57.3692}. 

Much work has been performed in the regime of intense ($\lesssim 10^{15}\;\textup{W}/\textup{cm}^2$) IR pulses. A recent comparison between experiment and theory showed perfect agreement \cite{PhysRevA.67.063405}. On the other hand, multiphoton ionization of rare gas atoms in the regime of high photon energies, which allows for direct ionization of one electron by a single photon, has been investigated at the free electron laser (FEL) facility at DESY in Hamburg which gave the first experimental evidence of a multi-photon process in the high photon energy regime \cite{PhysRevA.70.053401, PhysRevLett.94.023001}. Up to now, there still exist discrepancies between theory and experiment \cite{PhysRevA.77.023415} which may also be attributed to still limited experimental work in the field since the high-intensity VUV and XUV sources are presently under construction. Many predictions for further investigations already exist \cite{JPhysB.32.5629, EurophysLett.57.25, PhysRevLett.89.143401, PhysRevA.57.2832, JPhysB.34.L289, JPhysB.34.2245, JPhysB.32.L1, JSynchrotronRad.9.298, JPhysB.33.2287} but, to our knowledge, no information on the expected ADs of the photoelectrons is available. Since the ponderomotive potential is of the order of $\textup{meV}$ in this regime, ponderomotive scattering should play only a minor role. Additionally, the ionization happens directly into the continuum -- no channel switching due to ponderomotive shifting of resonant Rydberg states, affects the final electron state. Hence, the ADs should be dominated by the initial state of the atom and, with increasing laser intensity (and therefore higher electrical field strength), by rescattering effects with the parent ion. 

The aim of this paper is the investigation of (multiphoton) PADSs at high photon energies. It is organized as follows: In Section \ref{section:method} we explain our general theoretical approach to obtain angular distributions of photoelectrons which is followed by an explanation of the chosen model system in Section \ref{section:model_system}. Our results are presented in Section \ref{section:results} in detail for various laser parameters. The work ends with a discussion of the effects which contribute to the PADs and gives a physical explanation of the structures in terms of scattering processes by utilizing a simple analytical fit formula.

\end {section}

\begin{section}{Method}
\label{section:method}
\begin {figure}
\includegraphics[width=0.35\textwidth]{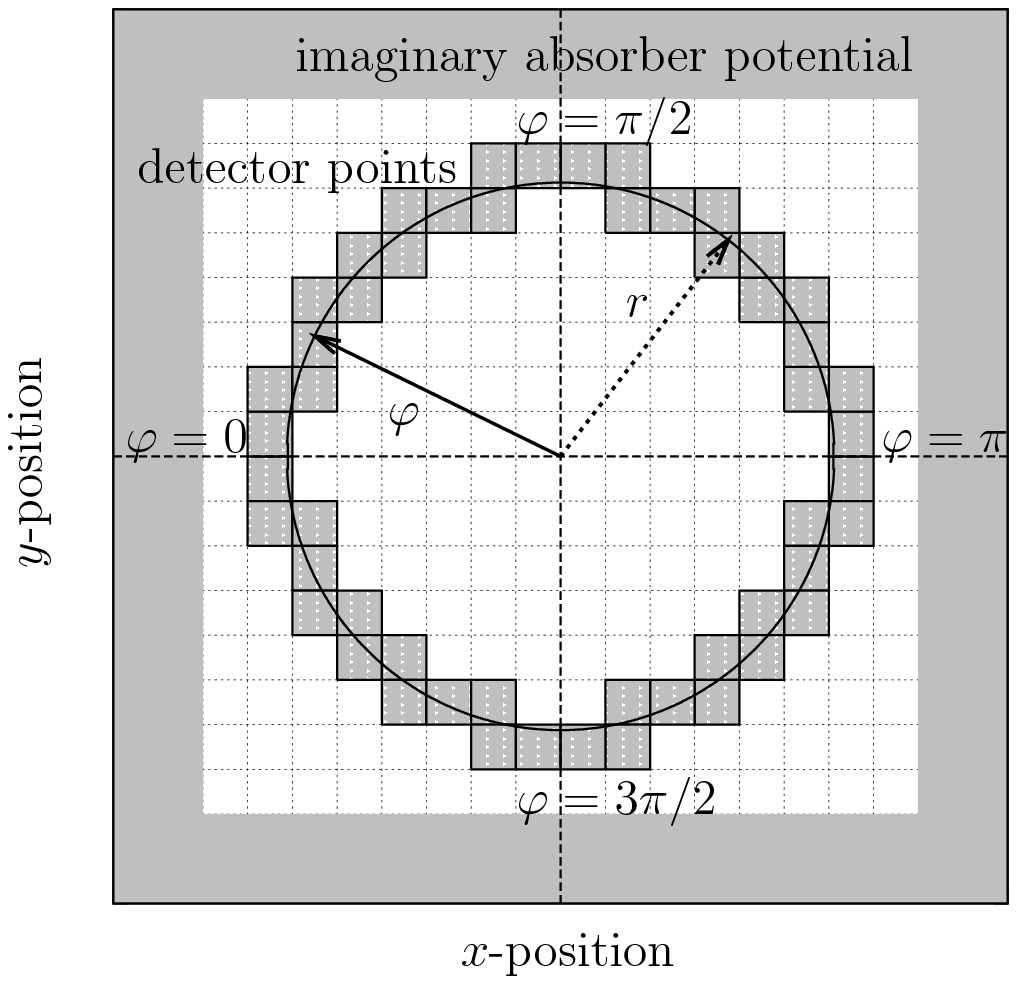}
\caption{Schematic drawing of the used detector implementation. The wave function is saved at a large fixed radius $r$ after which it is damped by  an imaginary absorbing potential. Typical detector parameters are $r\geq25.0 \;\textup{a.u}$ and more than $1500$ detector points. The atom is located in the center region at $r=0$.}
\label{fig:detector}
\end {figure}
We solve the two-dimensional time-dependent Schr\"odinger equation (TDSE) \cite{atomic_units}
\begin {equation}
 i \frac {\partial}{\partial t} \Psi(x,y,t) = \left(-\frac {\textup{d}^2}{2 \textup{d}x^2} - \frac {\textup{d}^2}{2\textup{d}y^2} + V(x,y,t) \right ) \Psi(x,y,t)
\end {equation}
for one particle on large spatial grids by means of an implementation of the \emph {Crank-Nicolson} procedure in combination with the operator-splitting technique which gives access to the solution of the two-dimensional problem. The potential term $V=V_{\textup{atom}}+V_{\textup{int}}$ is given by a time-independent part $V_{\textup{atom}}(x,y)$ and a time-dependent interaction part $V_{\textup{int}}(x,y,t)$ which describes the external laser field. 

The electro-magnetic field is treated classically within the \emph {dipole}-approximation. Then, the interacting part of the Hamiltonian for a  linearly polarized laser field along $x$-direction reads in length gauge 
\begin {equation}
 V_{\textup{int}}(x,y,t)= \exp\left(-\frac {(t-t_0)^2}{2\tau^2}\right) E_0 x\cos (\omega (t-t_0))\;,
\label{eq:v_laser}
\end {equation}
 where the envelope is assumed to be of Gaussian shape. $\tau$ describes the pulse duration (full width at half maximum FWHM), $t_0$ the time of maximum electrical field strength $E_0$ and $\omega$ is the photon energy.\\
 In order to minimize the computational grid sizes a dissipative absorbing potential with an imaginary part is included at the boundary of the grid. The decrease of normalization of the wave function due to absorption gives then an estimate of the total ionization rate. 
\begin {figure}
 \includegraphics[width=0.48\textwidth]{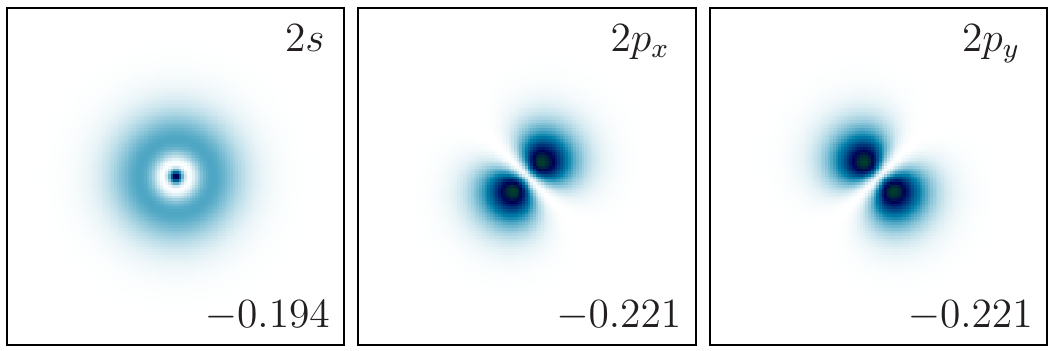}
 \caption {(color online) Density of the initial states of the regularized Coulomb potential, Eq.~\eqref{eq:reg_coulomb} with $\kappa=0.1$, in the $x-y$-plane calculated via TDSE propagation in imaginary time direction. The lower numbers give the energy eigenvalue of the state. The nomenclature of the states is chosen in analogy to the orbitals of the hydrogen atom.}
 \label{fig:initial_density}
\end {figure}
Special attention is paid to the calculation of energy spectra of the ionized electrons.  We implemented a realistic detector-like setup (cf. Fig.~\ref{fig:detector}) which allows for the calculation of energy- and angle-resolved photoemission spectra. The two-dimensional wave function $\Psi(t,r=\sqrt{x^2+y^2})$ is saved at the fixed radius $r$, where $r$ has to be chosen large enough to avoid near-field effects, for all time steps and hence the energy spectrum on the detector can be obtained by a Fourier transform with respect to the time $t$. After its detection the wave function is damped by the imaginary part of the potential, cf. Fig.~\ref{fig:detector}. We used $r\geq25.0$ and approximately $1500$ detector points allowing for a high angular resolution. During the data processing, the resolution is reduced to an angle element of $\Delta \varphi=2\pi/256$.

As initial conditions the eigenstates of the considered potential are used. They are obtained via propagation of the TDSE in imaginary time direction (time step $i\Delta t$) \cite{CompPhysComm.174.396} which is a procedure similar to self-consistent Hartree Fock calculations. Higher-lying states (above ground state) are constructed by an additional orthogonalization procedure at each imaginary time step. 

\end {section}
\begin{section}{Model system}
\label{section:model_system}

\begin {figure}
 \includegraphics[width=0.49\textwidth]{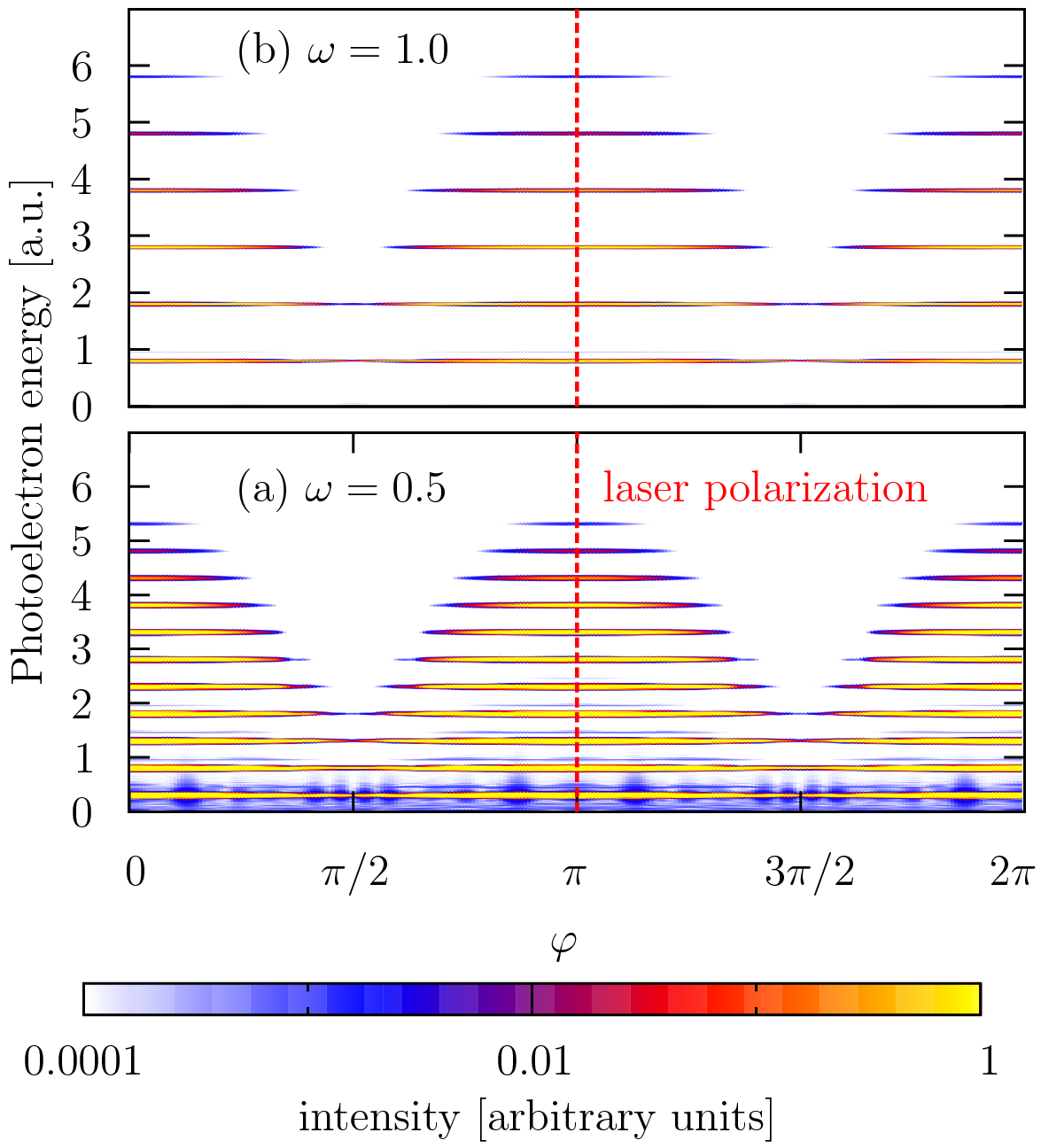}
 \caption{(color online) Angle- and energy-resolved spectrum of the ionization of the $2s$-state for two photon energies: (a) $\omega=0.5\;\textup{a.u.}$ and (b) $\omega=1.0\;\textup{a.u.}$ The electrical field strength is fixed at $E_0=0.1\;\textup{a.u.}$ corresponding to a laser intensity of $I=3.5\cdot 10^{14}\;\textup{W/cm}^2$. The direction of the linearly polarized laser field is indicated by the (red) dashed line.}
 \label{fig:photon_lines_s}
\end {figure}
\begin {figure}
 \includegraphics[width=0.49\textwidth]{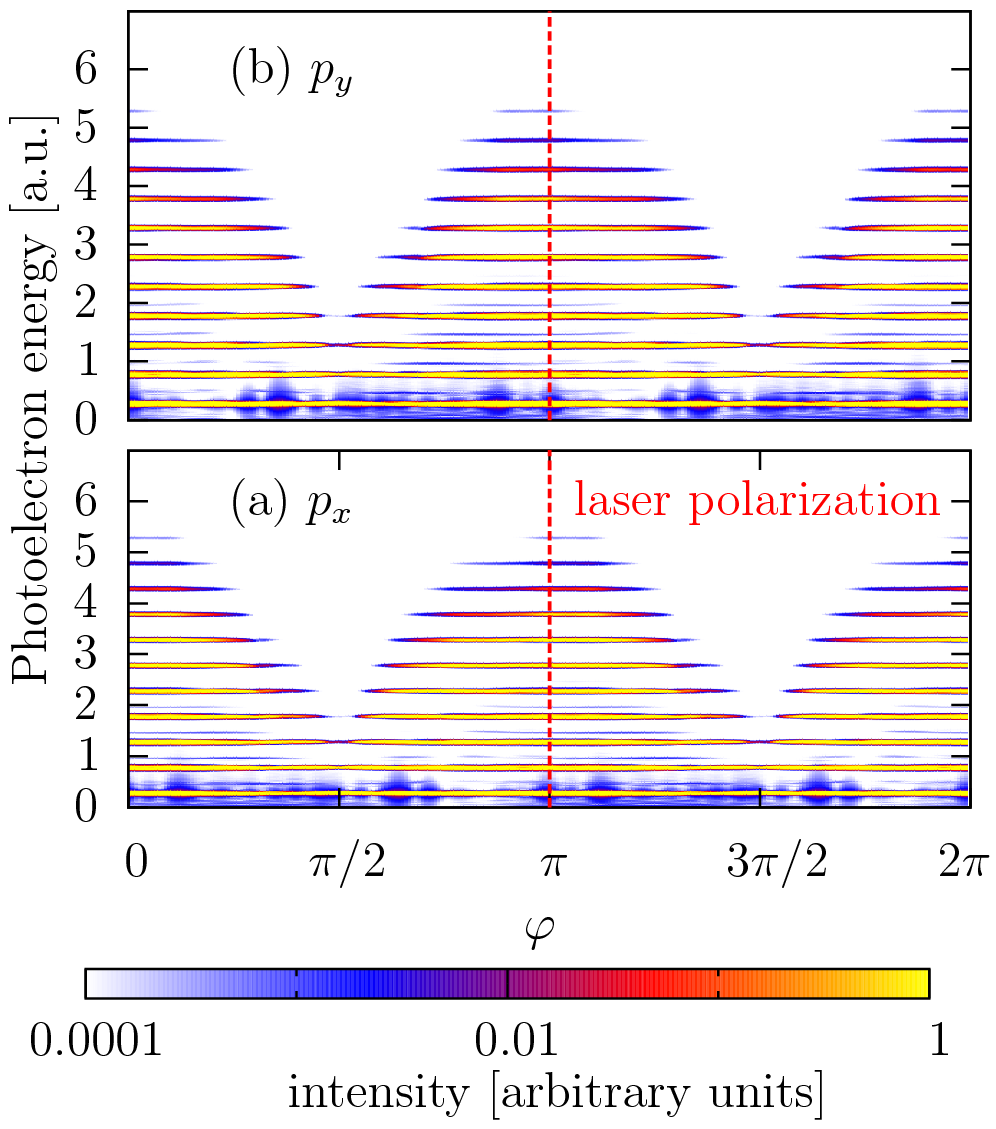}
\caption{(color online) The same as Fig.~\ref{fig:photon_lines_s} but for (a) the $p_x$- and (b) the $p_y$-state, excited with a  laser frequency of $\omega=0.5\;\textup{a.u.}$}
 \label{fig:photon_lines_p}
\end {figure}

The binding potential of the atom is represented by a two-dimensional regularized Coulomb potential
\begin {equation}
 V_{\textup{atom}}(x,y)=\frac {Z}{\sqrt{x^2+y^2+\kappa^2}} 
\label{eq:reg_coulomb}
\end {equation}
in our simulations. $\kappa$ is a small cut-off parameter preventing the singularity at $x=y=0$. This is necessary for the numerical treatment on the chosen cartesian grid. Furtheron, we investigate a hydrogen-like atom ($Z=-1.0\;\textup{a.u.}$ ) with a chosen  regularization of $\kappa=0.1$. Throughout this paper the ionization process of the $2s$ and the $2p_x$-like and $2p_y$-like states (cf. Fig.~\ref{fig:initial_density}) will be investigated. These chosen orbitals are typical for all $s$ and $p$ states. The orientation of the states with respect to the laser polarization is generated by the imaginary time propagation and is -- in some sense -- arbitrary. The explicit alignment of the states will be discussed lateron, see Section \ref{section:ionization_p}. The deeply-bound $1s$ ground state -- being present in the construction of eigenstates via the imaginary time propagation of the TDSE -- is non-physical due to the smoothing of the potential and is not considered for ionization in the following ($|E_{\textup{bind}}|>3 \;\textup{a.u.}$). Also its binding energy exceeds the photon energy several times, hence no or only small fractions could be ionized (the intensity of most of the laser fields considered is too small to allow for significant multiphoton ionization from this state). \\

The introduced artificial screening by the parameter $\kappa$ lifts the degeneracy of the $2p$-states and the $2s$-state, as can be seen in the printed eigenenergies in Fig.~\ref{fig:initial_density}. In analogy to the full hydrogen atom we will call the constructed states with $l=1$, for simplicity, $p_x$-state and $p_y$-state, although they are not aligned along the corresponding axis. In Section~\ref{section:p_angle_dep} the dependence of the angular distribution of photoelectrons on the explicit alignment of these states with respect to the laser polarization axis will be investigated in detail.

\end {section}

\begin {section} {Results}
\label{section:results}
\begin {figure*}
 \includegraphics[width=0.9\textwidth]{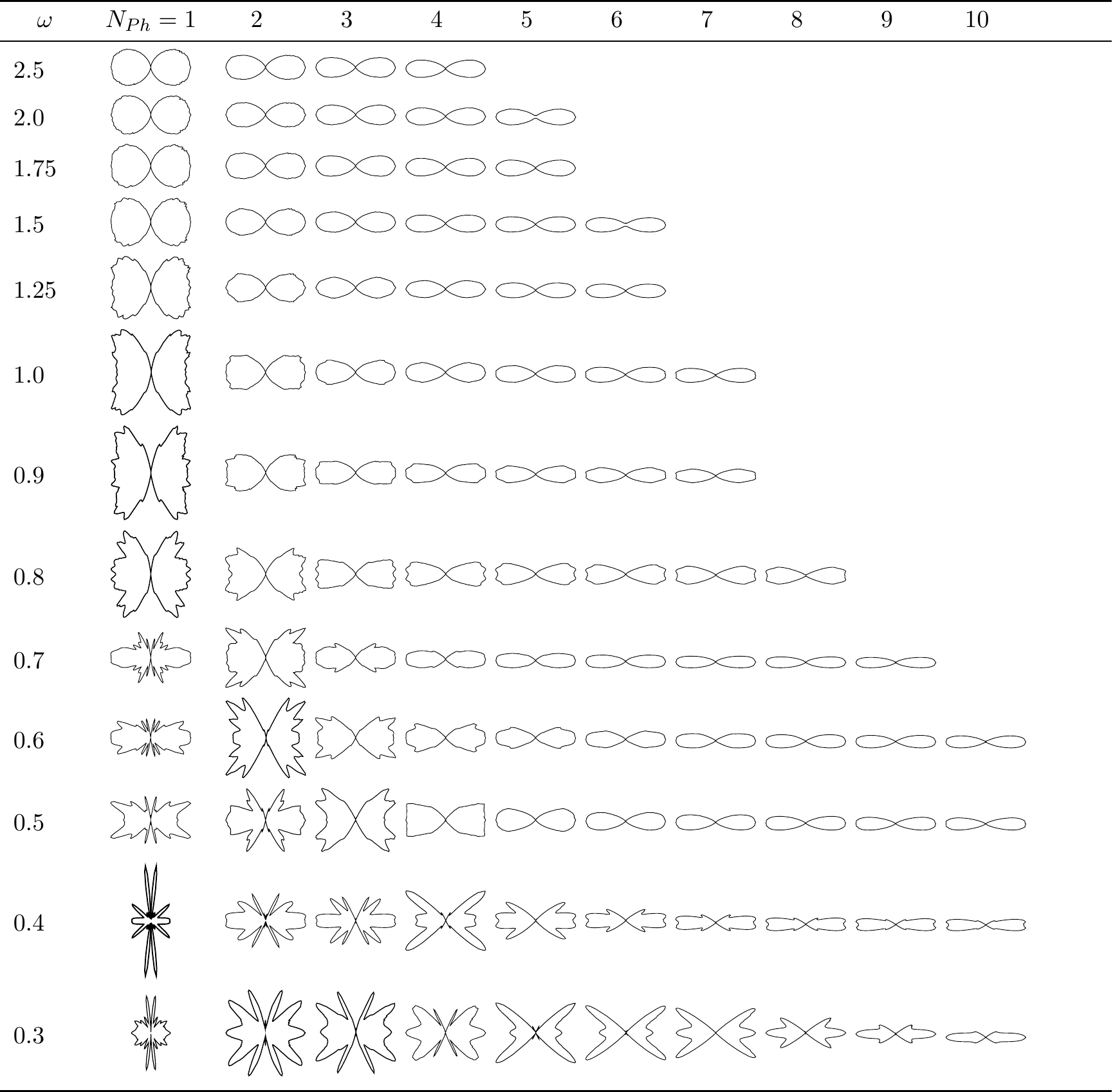}
  \caption{Angular dependence of photoelectrons for different photon orders $N_{Ph}$ (columns) and different photon energies $\omega$ (rows, given in atomic units) for the $2s$-state at an intensity of $I=3.5\cdot 10^{14}\;\textup{W/cm}^2$. The laser field is polarized in $x$-direction (horizontally).}
  \label{fig:w_dep_s}
\end {figure*}
\begin {figure*}
 \includegraphics[width=0.9\textwidth]{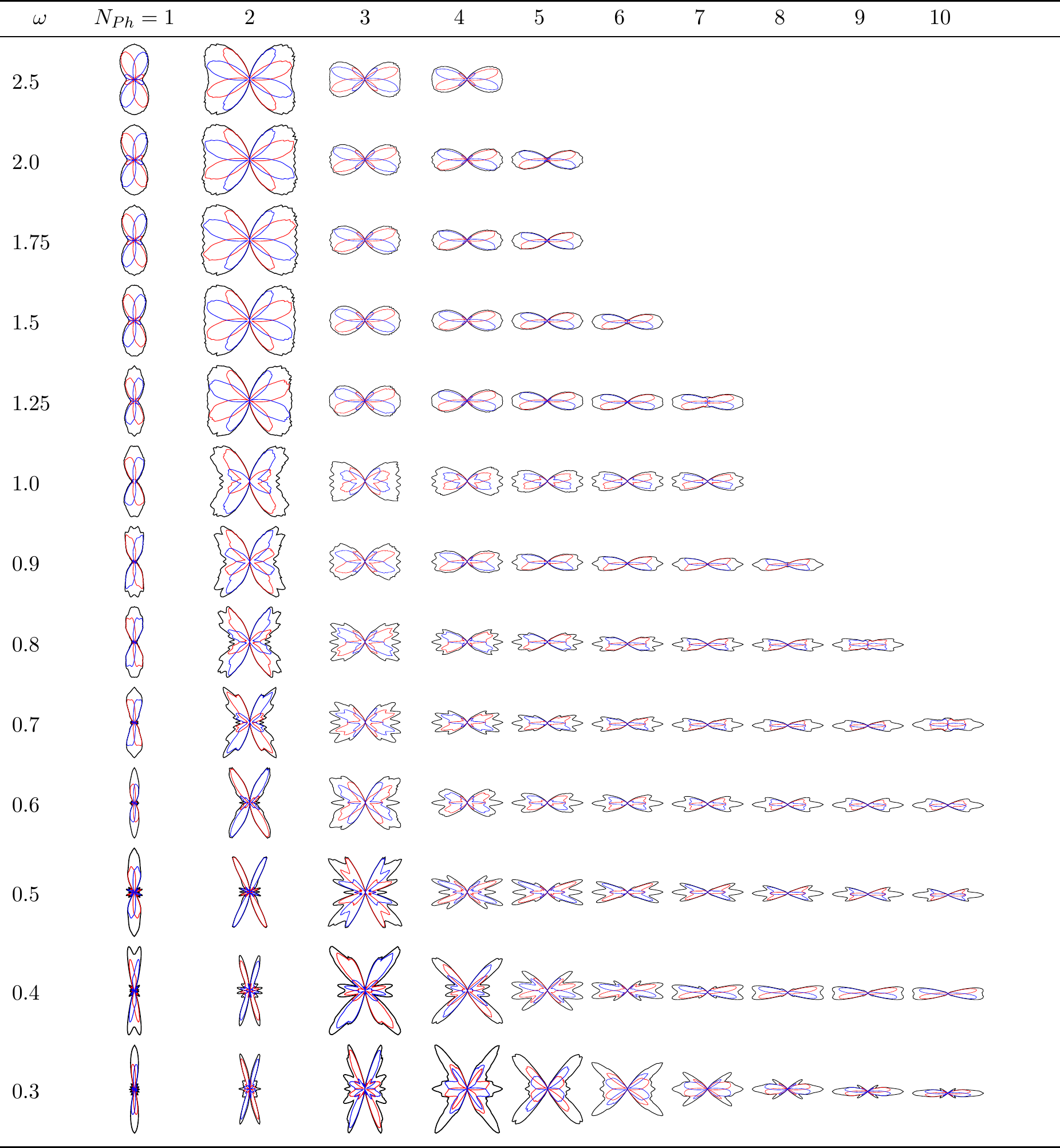}
  \caption{(color) Angular dependence of different photon orders $N_{Ph}$ (columns) for different photon energies $\omega$ (rows, given in atomic units) investigated for the $2p$-states at a laser intensity of $I=3.5\cdot 10^{14}\;\textup{W/cm}^2$. The colored lines indicate the angular distribution of photoelectrons of the single $2p_x$ (red lines) and $2p_y$ (blue lines) state. The laser field is polarized in $x$-direction (horizontally).}
  \label{fig:w_dep_p}
\end {figure*}
\begin {subsection}{Angular dependence of different photon lines for $s$-states}

We consider pulses of duration $\tau=30\;\textup{fs}$ to closely reproduce typical experimental situations with different photon energies and intensities. Since we use UV photon energies $\omega$ in our calculations the chosen duration should not affect the results significantly as many oscillations of the electrical field are present and we are far of the regime off few-cycle pulses.

Figs.~\ref{fig:photon_lines_s} and \ref{fig:photon_lines_p} show typical angle- and energy-resolved photoelectron spectra for the $s$ and the $p_x$/$p_y$-states. One clearly recognizes, for all cases, the formation of the ATI plateaus with distinct photon peaks separated by the photon energy $\omega$. The isolated photon lines can be identified for more than five orders of magnitude above the numerical noise level. 
In the case of the $s$-state two different photon energies are plotted for illustration: Fig.~\ref{fig:photon_lines_s} (b) shows the angular distribution for excitation with an energy of $27.2\;\textup{eV}$ and Fig.~\ref{fig:photon_lines_s} (a) the distribution of photoelectrons ionized with $13.6\;\textup{eV}$ radiation. The energetic cut-off in both spectra is approximately the same, which can be attributed to the corresponding small ponderomotive potential in both cases. This corresponds for the case a) of smaller photon energy, $\omega = 0.5\;\textup{a.u.}$, to twice as many photon peaks, compared to b).

A first inspection reveals that the photoelectron is ejected from the $s$-state predominantly along the polarization axis of the exciting field with a perfect angular symmetry around $\varphi=0$ and $\varphi = \pi$. In the case of both $p$-states one finds a small deviation from this symmetry, best seen in Fig.~\ref{fig:photon_lines_p} (a) and (b) at high photon orders whereas the main distribution --- at least for high photon orders --- is also ejected in directions near to the polarization axis of the electrical field.

For further investigation of the processes involved, it is useful to construct the angular dependence of each separate photon line from the full spectrum obtained numerically.  For this purpose we integrate the energy-resolved spectrum over an energy interval $E \dots E+\Delta E$ where $\Delta E$ is chosen according to the width of the photon line considered. Since no channel switching or high-lying Rydberg resonances are accessed (due to the direct ionization with one single photon in the lowest photon order) the substructures in the photon peaks are not of importance in our case and are integrated out. This procedure is performed separately for each photon energy and intensity for every occuring photon peak in the spectrum.

The angular dependence of photoelectrons initially bound in the isotropic $2s$-state is shown in Fig.~\ref{fig:w_dep_s} for different photon energies and photon orders. The columns correspond to the absorption of a fixed number of photons $N_{Ph}$ whereas the photon energy is varied for each row from $\omega=2.5\;\textup{a.u.}$ (top row) to $\omega=0.3\;\textup{a.u.}$ Each chosen energy is sufficient to ionize the electron directly by absorption of one single photon, cf. Fig.~\ref{fig:initial_density}. One clearly sees the increased absorption of additional photons in the continuum with decrease of the photon energy, as discussed above. For the case of $\omega=2.5\;\textup{a.u.}$ only four photon peaks can be identified in the spectrum whereas for $\omega=0.3\;\textup{a.u.}$ more than $n=15$ single photon peaks are found. The intensity of the laser pulse was chosen to be $I=3.5\cdot 10^{14}\;\textup{W/cm}^2$ for all photon energies. Therefore the ponderomotive potential $U_p=E_0^2/4\omega^2$ increases from top ($U_p=0.0004\;\textup{a.u.}$) to bottom ($U_p=0.028\;\textup{a.u.}$) but it is small compared to the binding energy and the photon energy of the laser field. Therefore, all higher-order ionization processes ($N_{Ph} >1$) are due to multiphoton absorption. Tunneling effects are expected to play only a negligible role.
\end {subsection}

\begin {subsection} {Angular dependence of different photon lines for $p$-states}
\label{section:ionization_p}

\label{section:p_angle_dep}
To investigate states with different angular momentum, as e.g., the $2p$-states with $l=1$, one has to keep in mind the orientation of the orbital with respect to the laser polarization axis. To construct the angular dependence of the photoelectrons for randomly aligned states it is necessary to address the specific orientational effects a single orbital shows in its PADs. In order to use the methods developed above we combine the photoelectron spectra of different states, i.e. for our special case (of the $p$-subspace) both $2p_x$ and $2p_y$-states. 
Since the imaginary time stepping method constructs an orthogonal set of eigenstates, which are arbitrarily aligned with respect to the laser field, the dependence of the PAD on the explicit orientation of the $2p_x$ and $2p_y$ states has to be investigated. For this purpose we consider, in addition, two rotated, orthogonal states $2p_x'$ and $2p_y'$ (cf. Fig.~\ref{fig:cf_pxpx}). Technically, the rotation of the states obtained numerically is performed by superposition of both original states,  $2p_x$ and $2p_y$, in such way that the final orientation of the new states is rotated by $45^\circ$ compared to the original orientation. \\

The obtained PADs for both cases are given in Fig.~\ref{fig:cf_pxpx}. The combined intensity of both states, $I(p_x') + I(p_y')$,  shows exactly the same behavior as the intensity of the non-rotated states  $I(p_x)+I(p_y)$. Therefore, special orientation-dependent features in the total angular distribution of photoelectrons can be excluded.\\
Knowing this, it is sufficient to calculate the PADs of the $p_x$ and $p_y$ states. Since for atoms in the gas phase the orbitals occur randomly oriented, for every given state its orthogonal complement can also be found. We assume here, that the electrical field of the exciting laser is changing rapidly enough (high photon energies), such that the dipole moment of the atomic orbital does not play an important role and the atom is not aligned in some special way with respect to the laser polarization axis. Because the total PAD of each pair has the same shape, it is only necessary to know the ionization behavior of one single pair. In the following, we will therefore consider only two orthogonal states for the investigation of the total angular distributions of photoelectrons for $p$-states.\\

\begin {figure}
  \centering{ \hspace{1.25cm}\includegraphics[width=0.32\textwidth]{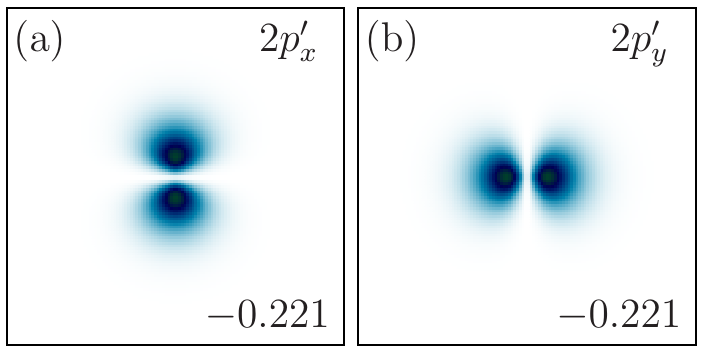}}
  \centering{ \includegraphics[width=0.49\textwidth]{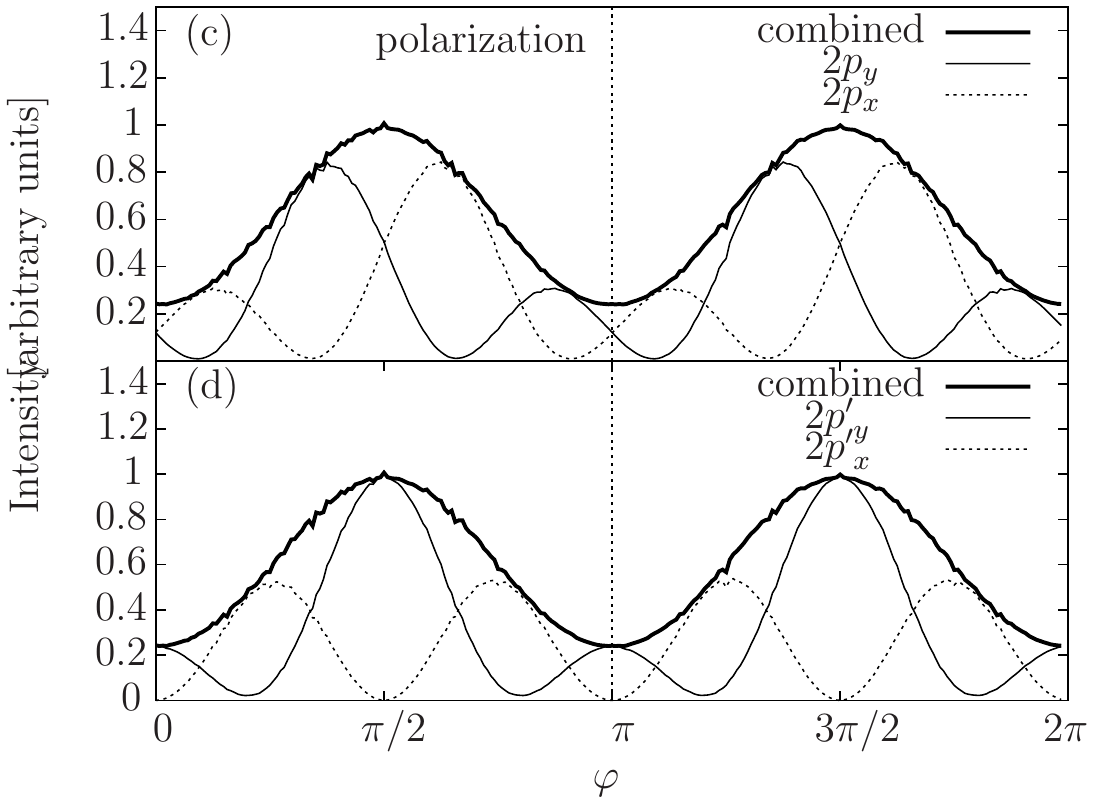}}
 \caption {Comparison of the photoelectron angular distributions for different orientations of the initial states. The upper figures give the densities of the rotated states (a) $2p_x'$ and (b) $2p_y'$. The PADs of the non-rotated $2p$-states, cf. Fig.~\ref{fig:initial_density}, are shown in (c) and for the rotated states in (d). The thin lines indicate the data for each individual orientation whereas the bold line displays the total intensity $I(2p_x)+I(2p_y)$ in (c) and  $I(2p_x')+I(2p_y')$ in (d), respectively. }
 \label {fig:cf_pxpx}
\end {figure}
\begin {figure*}
 \includegraphics[width=0.9\textwidth]{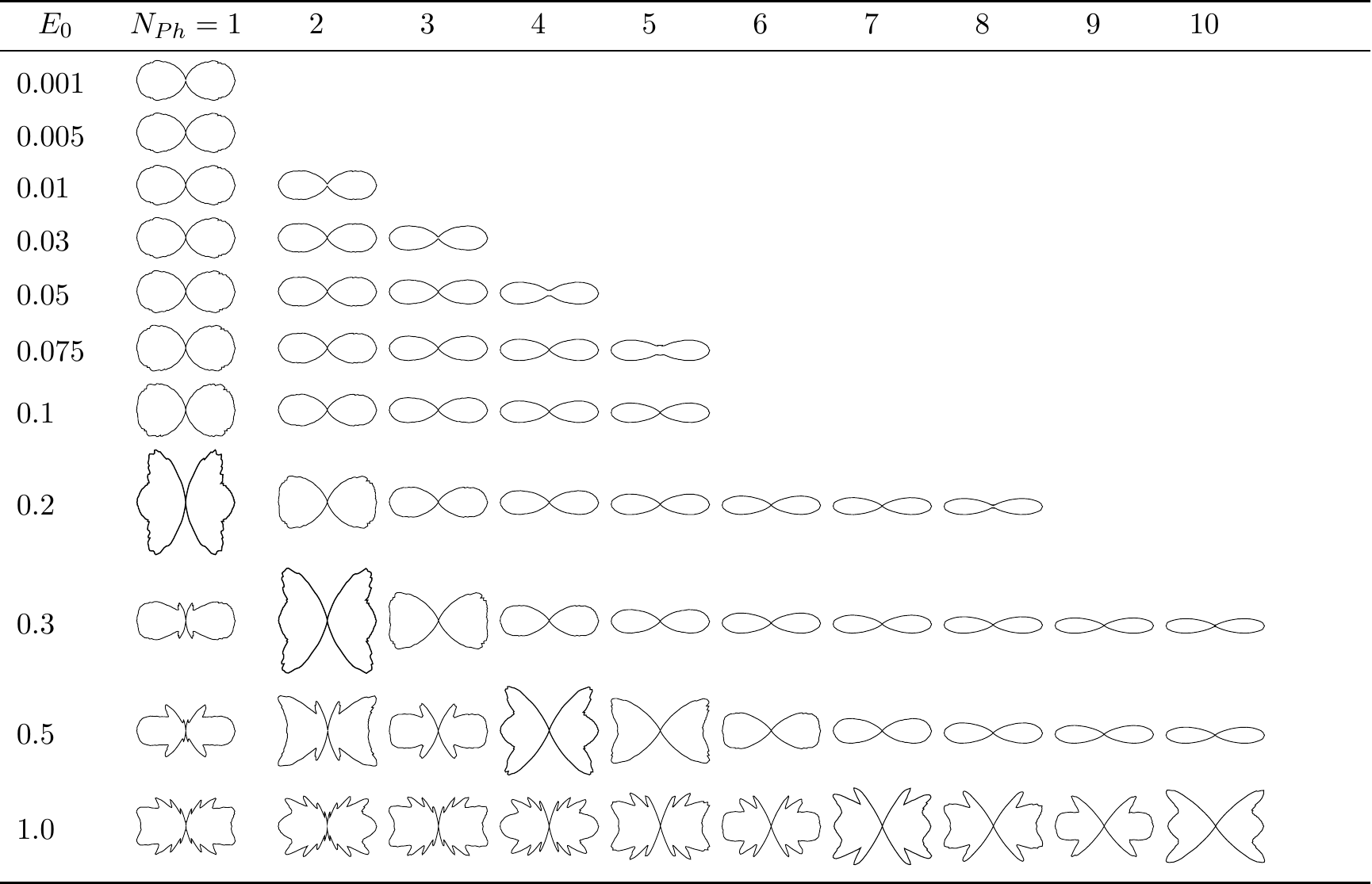}
  \caption{Photoelectron angular dependence for different photon orders $N_{Ph}$ (columns) and different intensities (rows, electrical field amplitude $E_0$ given in atomic units) for the $2s$-state at a fixed photon energy of $\omega=1.69\;\textup{a.u.}$ The laser field is polarized along the $x$-direction (horizontally).}
  \label{fig:e_dep_s}
\end {figure*}
Fig.~\ref{fig:w_dep_p} shows the obtained PADs from our TDSE simulation in dependence on the photon energy (rows) and the photon order (column) for the $2p$-state, similar to the case of the $2s$-state, Fig.~\ref{fig:w_dep_s}. But for this case, as discussed above, each angular distribution contains three different graphs. The oriented $p_x$-state (blue line), the oriented $p_y$-state (red line) and the total intensity, plotted as an envelope (black line). As in the $2s$ case, one easily recognizes the emergence of additional photon lines with decreasing photon energy and the occurence of richly structured PADs. 

Additionally, the appearing of three regimes corresponding to different orientation of the PADs with respect to the laser polarization axis can be pointed out: in the first photon order ($N_{Ph}=1$) the maximum intensity for ejected photoelectrons is oriented perpendicular to the laser polarization axis (left-most column in Fig.~\ref{fig:w_dep_p}). In the third regime (large numbers of absorbed photons) the PAD is aligned along the field polarization axis (right columns in Fig.~\ref{fig:w_dep_p}). Finally, in the intermediate region, a transition-like behavior is observed which is characterized by complex angular modulations.

\end {subsection}

\begin {subsection} {Intensity dependence of PADs}

Since scattering effects should become more prominent at higher intensities, or vice versa, decrease for low intensities, we calculated the PADs for different intensities of the laser field. The results are shown in Fig.~\ref{fig:e_dep_s} for a fixed photon energy $\omega=1.69\;\textup{a.u.}$. The columns again indicate the absorption of single photons and the rows mark different intensities from low (top, $I = 3.5\cdot 10^{10} \;\textup{W}/\textup{cm}^2$ ) to high (bottom, $I=3.5\cdot 10^{16}\;\textup {W}/\textup{cm}^2$) \cite{intensity_comment}. Again, with variation of the intensity, the number of observed photon orders in the spectrum varies: from linear excitation, by absorption of one single photon with perfect dipole-like shape (top-most row in Fig.~\ref{fig:e_dep_s}) in the perturbative regime, to the absorption of many photons with a richly structured angular dependence of the photoelectrons for different photon orders (lower rows in Fig,~\ref{fig:e_dep_s}), indicating scattering effects of the ejected electron with its parent ion. This trend is typical for all initial states ($2s$ and $2p$ states).

\end {subsection}

\begin {subsection} {Physical explanation of the PADs in terms of scattering processes}
\label {section:explanation}
First of all, the shapes of the angular distributions have a qualitatively similar structure as found in experiments for IR photon energies \cite{JPhysB.31.4617}. But for all cases considered within the present work  (except for very high-intensity excitations shown in Fig.~\ref{fig:e_dep_s}, lowest row) the well-known ATI plateaus with characteristic energies of $2.5 U_p$, $4.5 U_p$ and $10 U_p$ are energetically located below the first observed photon peak, since the ponderomotive energy is small, $U_p < 0.03\;\textup{a.u.}$.  Therefore, the structures in the angular distribution cannot be attributed to the same effects as in the IR photon case, i.e., to the onset of the ATI-plateau \cite{JPhysB.31.4617,JPhysB.27.L703}.

Nevertheless, scattering mechanisms should play the dominant role in the formation of the patterns observed. This we conclude from the fact that the richly structured shapes of the PADs, Figs. \ref{fig:w_dep_s}, \ref{fig:w_dep_p} and \ref{fig:e_dep_s}, disappear both, for very large photon energies, and for simultaneous absorption of many photons at low intensities, respectively, where the electron is lifted high into the continuum and the absorbed field energy is converted directly into translational motion (kinetic energy) away from the atom. In this case, it is unlikely that the electric field, which changes very rapidly, drives the electron back to undergo a rescattering process. An analogous argument holds for the case of low intensities. The electrical field of the laser is not strong enough to modify the path of the fast-travelling electronic wave packet. In the following we will examine the obtained PADs more in detail to isolate traces of such scattering events. \\

The shape of the PAD for the multiphoton ionization of order $n$ is often fitted to a sum of even Legendre polynomials $P_{2l}$ \cite{Compton_1984}
\begin {equation}
     I^{(n)}(\theta)=\frac{\sigma}{4\pi}\left(1+\sum_{l=1} ^ n \beta^{(n)}_{2l} P_{2l} (\cos \theta)\right)\;,
\label{eq:legendre_sum}
\end {equation}
to correctly describe the angular momentum the outgoing electron wave carries where $\theta$ is the angle between the electron and the field polarization axis.
In the case of the absorption of one single photon Eq.~\eqref{eq:legendre_sum} transforms into the well-known formula of Cooper-Zare \cite{starace1982},
\begin {equation}	
   I(\theta)=\frac{\sigma}{4\pi}\left[1+ \beta P_2(\cos \theta)\right ],
\label{eq:cooper_zare}
\end {equation}
with the dipole-anisotropy parameter $\beta$ and the angle-integrated ionization cross section $\sigma$.
Eq.~\eqref{eq:cooper_zare} describes the PAD for the ionization of randomly oriented atomic or molecular systems by a linearly polarized laser field.  Its shape is, therefore, completely described by the two quantities $\beta$ and $\sigma$.

If we now assume that, in the other regimes, scattering effects, and therefore changes of the angular momentum of the electron,  influence the PADs, additional terms in Eq.~\eqref{eq:legendre_sum} beyond the first order, Eq.~\eqref{eq:cooper_zare}, are needed. To this end, we fit a sum of Legendre polynomials $P_{2l}$ with $l$ up to four via a least square fitting routine to our simulation data.

 We will concentrate our discussion on the $2s$ state. In principle, the same results and arguments hold for the $2p_x$- and $2p_y$-states. The fitting results, i.~e., the first four fit parameters $\beta_{2 \dots 8}$, in dependence on the photon energy and the intensity  for the first two photon lines are given in Fig.~\ref{fig:beta_params_s_omega} and Fig.~\ref{fig:beta_params_s_int}. Higher-order processes would involve additional polynomials in Eq.~\eqref{eq:legendre_sum} making the fitting procedure less reliable.
\\
\begin{figure}
 \includegraphics[width=0.49\textwidth]{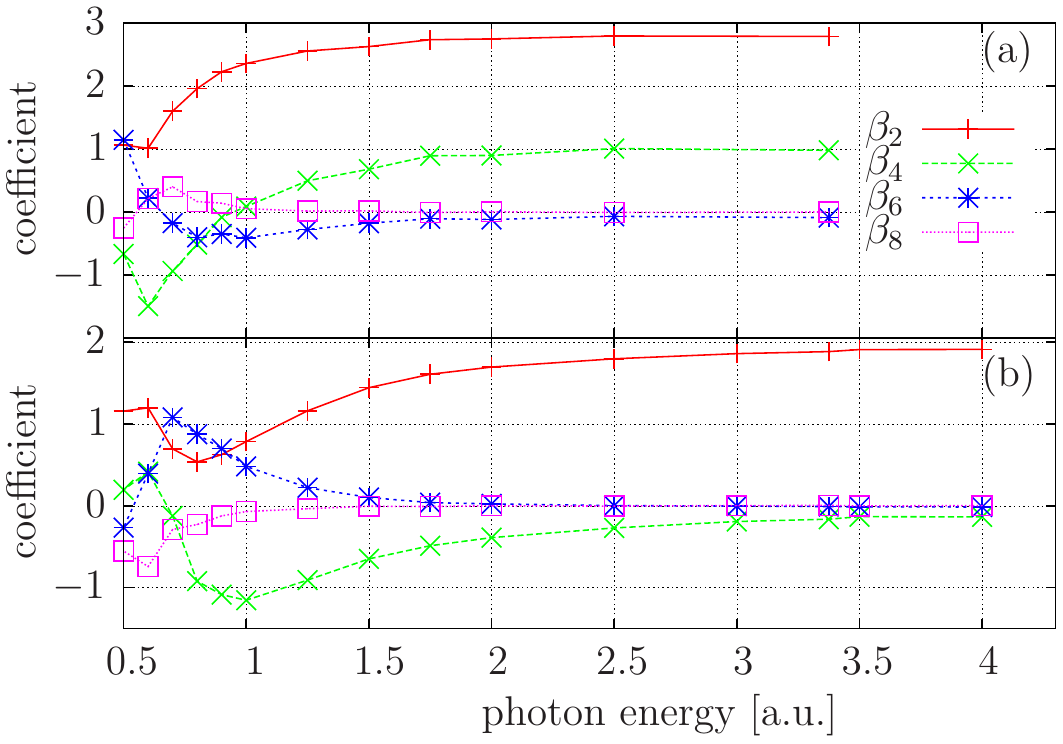}
  \caption {(color online) Parameters $\beta_l$ of Eq.~\eqref{eq:legendre_sum} calculated via least square fit to our numerical data for the first (b) resp. second (a) photon peak of the PADs of the  $2s$-state vs. the photon energy $\omega$ for a fixed intensity of $I=3.5 \cdot 10^{10}\;\textup{W/cm}^2$.}
 \label{fig:beta_params_s_omega}
\end{figure}

\begin{figure}
 \includegraphics[width=0.49\textwidth]{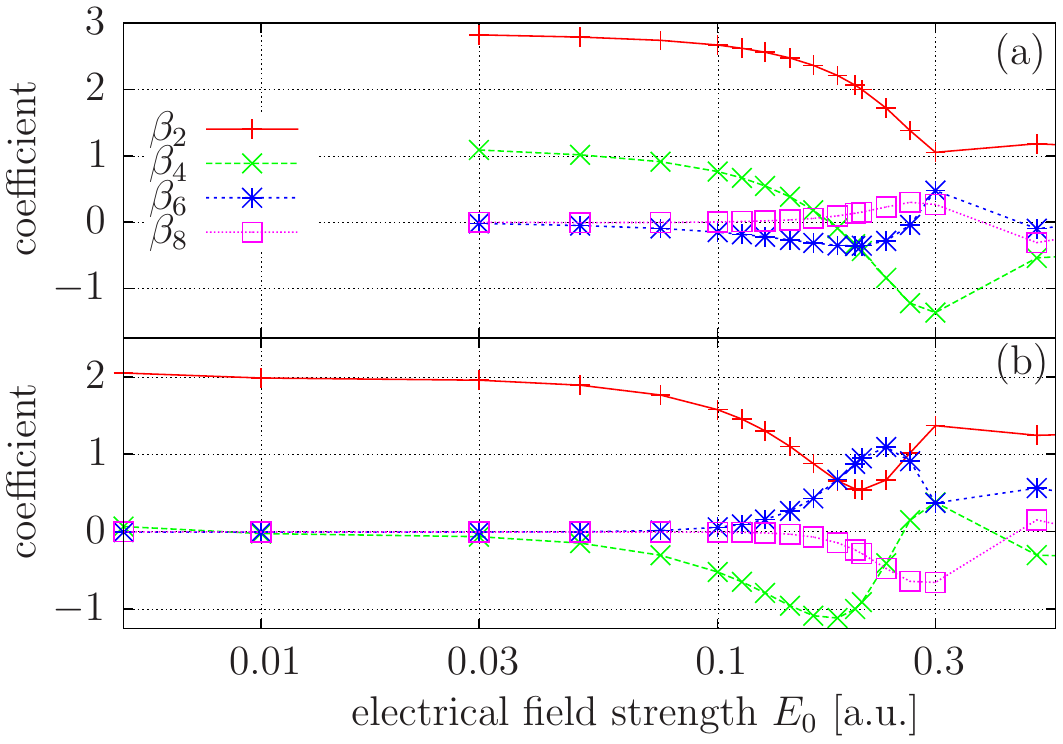}
  \caption {(color online) Intensity dependence of $\beta_l$ parameters in Eq.~\eqref{eq:legendre_sum} for the $2s$-state at a fixed photon energy of $\omega=1.69\;\textup{a.u.}$ Again (b) gives the result for the first photon peak (single-photon absorption) and (a) the data for the second photon line.}
 \label{fig:beta_params_s_int}
\end{figure}

\begin {figure}
 \includegraphics[width=0.49\textwidth]{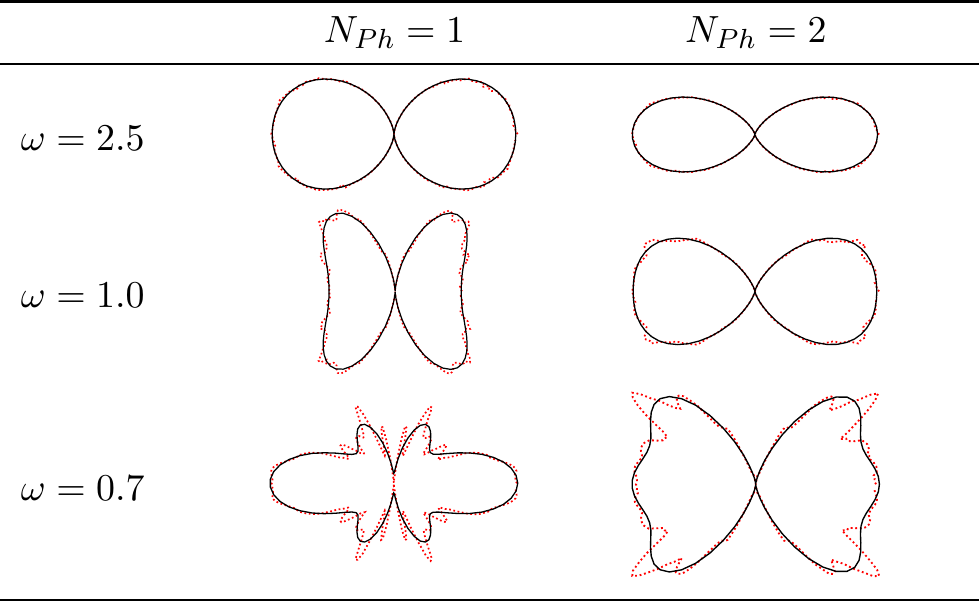}
  \caption{(color online) Comparison of PAD fitted according to  Eq.~\eqref{eq:legendre_sum}  (black solid lines), with the accurate numerical data ([red] dashed lines) for some selected typical cases of $\omega$, see Fig.~\ref{fig:w_dep_s}.}
  \label{fig:cf_fit}
\end {figure}

First, we will discuss the behavior for the limit of large photon energies or low intensities. This is the regime where scattering effects are expected to play only a minor role in the angular distributions of photoelectrons. And indeed, the intensity distribution follows the Cooper-Zare formula,  Eq.~\eqref{eq:cooper_zare}, in the case of single-photon absorption [Fig.~\ref{fig:beta_params_s_omega} (b) and Fig.~\ref{fig:beta_params_s_int} (b)] with an anisotropy parameter of $\beta_2=2.0$. This corresponds exactly to the theoretical value for dipole radiation. All higher-order parameters $\beta_{l>2}$ vanish. Analogously, the first two parameters $\beta_2$ and $\beta_4$ are required to describe the distribution in the case of the second photon line, as it is expected from Eq.~\eqref{eq:legendre_sum} for $n=2$. 

If the photon energy is decreased, the initial dipole shape gets lost and the coeffecients of higher-order polynomials, $\beta_4, \beta_6, \dots$, increase where the next higher-order contribution is filled one after another. This behavior becomes most obvious in the case of the first photon line, cf. Fig.~\ref{fig:beta_params_s_omega} (b), where the $\beta_2$ parameter decreases from its dipole value $\beta_2=2.0$, and the magnitude of the term $\beta_4$ is becoming larger ($\omega \approx 2.5 \dots 1.5 \; \textup{a.u.}$). In the following step ($\omega \lesssim 1.5\; \textup{a.u.}$), the third-order parameter $\beta_6$ gets involved to describe the PAD correctly.
This behavior can be spotted directly in the shape of the PADs, cf. Fig.~\ref{fig:w_dep_s}, first column $N_{Ph}=1$, where up to a value of $\omega \approx 1.5\;\textup{a.u.}$ the dipole shape dominates and only small deviations are present. With decreasing photon energy the PADs become compressed along the $x$-direction and some parts of the electrons are scattered in different directions giving at low photon energies finally rise to the observed complex structured PADs. 

 Accordingly, due to the higher kinetic energy of the photoelectrons being ionized by the simultaneous absorption of two photons and the corresponding smaller possibility of rescattering, the occupation of higher-order terms in Eq.~\eqref{eq:legendre_sum} sets in at smaller photon energies as can be seen in direct comparison of the (red) lines for $\beta_2$ in Fig.~\ref{fig:beta_params_s_omega} (b) and (a). For even lower photon energies, $\omega<0.7\;\textup{a.u.}$ the angular distribution becomes very irregular, cf. Fig.~\ref{fig:w_dep_s} lowest rows, and the fitting procedure fails.\\

 The same behavior can be observed by increasing the intensity of the laser field, cf. Fig.~\ref{fig:beta_params_s_int}, where also the higher-order processes are accessed one after another. As in the previously mentioned case of variation of the photon energy, the modifications can be directly spotted in the PADs, cf. Fig.~\ref{fig:e_dep_s}. Again, the dipole-like shape at low intensities is, with increase of the laser intensity, compressed along the polarization axis and parts of higher-order processes become relevant. This observation is condensed in the fact that higher-order polynomials in Eq.~\eqref{eq:legendre_sum} are accessed ($\beta_4$ and $\beta_6$ in Fig.~\ref{fig:beta_params_s_int}). 

 This lets us conclude, that scattering processes and, therefore, a modification of the angular momentum of the outgoing electron with the corresponding different angular dependencies, are dominant and responsible for the complex photoelectron angular distributions. The richly structured PADs are, therefore, also in the case of high photon energies at high intensities (but small ponderomotive forces), a consequence of the scattering processes the electron undergoes on the ion on its way from its creation by (multi) photon ionization to the detector.\\

A final remark shall be made on the success of the fitting procedure, cf. Fig.~\ref{fig:cf_fit}. Given the complex structure of the PADs, it is remarkable that the fit of only four Legendre polynomials is sufficient for resolving the main contributions to the PAD, at least for photon energies larger than $\omega=0.5\;\textup{a.u.}$ For large photon energies ($\omega=2.5\;\textup{a.u.}$) the shape of the PAD is completely describable by such a series, the (red) dashed line of the numerical data and the black solid line of the fitted function show almost no deviations for both photon orders under investigation. In the transition-like region, in Fig.~\ref{fig:cf_fit} represented by the case of $\omega=1.0\;\textup{a.u.}$, first small substructures can be spotted, being not captured by the fitted polynomial. But the main contributions are still resolved within high accuracy. In the last case of $\omega=0.7\;\textup{a.u.}$ given, the formation of jet-like structures at angles of approximately $60^\circ$ off the polarization axis of the laser plays against the fitting procedure, but still, the main contributions are accounted for. For even smaller photon energies, the strongly peaked, very sharp jet-like structures, cf. Fig.~\ref{fig:w_dep_s}, require to  extend the expansion, Eq.~\eqref{eq:legendre_sum} to orders higher than $l=4$. 

\end {subsection}
\end {section}

\begin{section}{Conclusions and Outlook}
\label{section:discussion}
In this paper we have investigated the angular distributions of photoelectrons being excited by linearly polarized laser fields at high photon energies which are (or will soon be) available at free electron laser facilities. The observed PADs are, depending on the photon energy $\omega$ and the electrical field $E$ of the laser, richly structured showing  prominent side lobes and jets which were observed before in the IR regime.  But, due to the small ponderomotive energy, the mechanisms cannot be attributed to the well-known formation of ATI plateaus. Nevertheless we have provided clear evidence that these effects are caused by scattering of the ionized electron on the ion when it is driven back by the laser field. This has been shown in terms of the analytical fit formula, Eq.~\eqref{eq:legendre_sum}, where subsequently higher-order terms are needed as the possibility of scattering events increases.

 While our results where obtained for a 2D model atom, we expect that the main features will survive in the 3D case. The present results are of relevance for single atoms and molecules (see e.g. \cite{prl.100.203003}). They also apply to atom ensembles because, as we have shown, the total PAD of electrons ionized from all $p$-states is independent of the orientation of the atoms relative to the laser field. The high sensitivity of the PAD to the field strength allows for accurate detection of the laser intensity at the place of the atoms. Furthermore, the PAD in the UV and XUV regimes will be of high interest to pump-probe experiments with femtosecond to sub-femtosecond time resolution, e.g. \cite{PhysRevLett.99.247601,mauritsson:073003}. Finally, it will be of high interest to extend the present analysis to many-electron atoms by including correlation effects. This can be done by using e.g. quantum kinetic equations or non-equilibrium Green's function methods \cite{dahlen07,kwong98,haberland01,kremp99} which is subject of ongoing work.

\end {section}

\begin{acknowledgments}
The authors acknowledge stimulating discussions with L. Kipp. This project has been supported by the \emph{Innovationsfond of the state Schleswig-Holstein}. 
\end{acknowledgments}

\bibliographystyle {apsrev}

\begin{thebibliography}{38}
\bibitem{atomic_units} Within this work atomic units (a.u.) are used ($\hbar=e=m_e=1.0$).
\bibitem{intensity_comment}The highest intensities are not yet available in current experimental situations at the high photon energies considered. Also, the Keldysh parameter becomes larger than $1$ for $E_0=1.0\;\textup{a.u.}$ and the main ionization mechanism switches from MP ionization to tunnel ionization.
\end{thebibliography}

\end{document}